\documentclass[aps,prl,reprint,superscriptaddress,amsmath,amssymb,floatfix]{revtex4-2}
\usepackage[T1]{fontenc}
\usepackage{mathptmx}
\DeclareSymbolFont{radialletters}{OML}{lmm}{m}{it}
\DeclareMathSymbol{\varrho}{\mathord}{radialletters}{"25}
\DeclareMathAlphabet{\mathcal}{OMS}{cmsy}{m}{n}
\usepackage{graphicx,bm}
\usepackage[hidelinks]{hyperref}
\hypersetup{pdftitle={Rotational Asymmetry beyond Phase Rigidity in Non-Hermitian Eigenfunctions},pdfauthor={Kyu-Won Park, Soojoon Lee, Kabgyun Jeong},pdfsubject={Final manuscript}}
\newcommand{\dd}{\mathrm d}
\newcommand{\rOm}{r_{\Omega}}
\newcommand{\smpage}[1]{\clearpage\thispagestyle{empty}\vspace*{-\dimexpr1in+\voffset+\topmargin+\headheight+\headsep+\topskip+1pt\relax}\noindent\hspace*{-\dimexpr1in+\hoffset+\oddsidemargin\relax}\makebox[0pt][l]{\includegraphics[width=\paperwidth,height=\paperheight,page=#1]{Supplement.pdf}}}
\begin{document}
\lefthyphenmin=2\righthyphenmin=3
\title{Rotational Asymmetry beyond Phase Rigidity in Non-Hermitian Eigenfunctions}
\author{Kyu-Won Park}
\affiliation{Department of Mathematics and Research Institute for Basic Sciences, Kyung Hee University, Seoul, 02447, Korea}
\author{Soojoon Lee}
\affiliation{Department of Mathematics and Research Institute for Basic Sciences, Kyung Hee University, Seoul, 02447, Korea}
\affiliation{School of Computational Sciences, Korea Institute for Advanced Study, Seoul 02455, Korea}
\author{Kabgyun Jeong}
\email{kgjeong6@snu.ac.kr}
\affiliation{Institute of Computer Technology, College of Engineering, Seoul National University, Seoul 08826, Korea}
\affiliation{School of Computational Sciences, Korea Institute for Advanced Study, Seoul 02455, Korea}
\date{\today}
\begin{abstract}
Phase rigidity, the standard measure of eigenvector nonorthogonality, vanishes at an exceptional point (EP). We show that this collapse need not imply phase spreading and can mask persistent phase organization. Using quantum and classical relative entropies linked by the Holevo bound and the phase--Fourier entropic uncertainty relation, we uncover rotational asymmetry within this organization. In both a classical microcavity and quantum hydrogen, phase entropies peak and Fourier entropies dip on a broad parameter scale. Near the EP, one branch gains rotational asymmetry while the other loses it despite common rigidity loss. Changes in quantum rotational asymmetry dominate those in classical asymmetry, consistent with the weak-pattern scaling of the uncertainty gap. This hidden phase organization predicts testable probe-induced splitting axes relevant to EP sensing and allows distinct rotational asymmetries at fixed geometric imaginarity. Rotational asymmetry thus reveals non-Hermitian eigenfunction structure beyond nonorthogonality.
\end{abstract}
\maketitle
\raggedbottom
\setlength{\parskip}{0pt}

Open classical and quantum systems support resonances with complex eigenvalues and nonorthogonal eigenfunctions. Exceptional points (EPs), where both coalesce, expose the coupled singular structure of spectra and modes \cite{Kato1966,AshidaGongUeda2020}. Spectrally, their branch points produce nonanalytic resonance splitting under generic perturbations \cite{Wiersig2023,Schomerus2024}. The eigenvalues retain a direct physical interpretation: their real parts give frequencies or energies, whereas their imaginary parts determine decay rates. Eigenfunction nonorthogonality affects perturbation sensitivity \cite{Wiersig2023,Schomerus2024}, laser excess noise through the Petermann factor \cite{Wiersig2023,Wang2020}, and transient amplification \cite{Makris2014}. It also enters quantum information flow \cite{Kawabata2017} and entanglement \cite{Lee2022Exceptional,Chang2020Entanglement}. Despite these roles, how nonorthogonality is expressed in an eigenfunction's internal structure remains unclear.

A transpose-symmetric Hermitian reference admits real eigenfunctions, whose local phases are locked to $0$ or $\pi$ up to a global phase. Opening can release this locking and increase nonorthogonality while preserving similar intensities \cite{Jeong2020}. In a reciprocal complex-symmetric representation, phase rigidity measures this nonorthogonality through the normalized left--right overlap \cite{Wiersig2023,Bulgakov2006,Lee2008Petermann,Brouwer2003}. Phase contributions from different regions can cancel in this overlap even when local phases remain clustered around distinct preferred directions. Falling rigidity therefore does not determine whether the phase distribution broadens or sharpens as the EP is approached. Resolving this hidden structure calls for a distributional framework embedding phase rigidity.

Entropic uncertainty constrains simultaneous concentration in complementary representations \cite{MaassenUffink1988,BialynickiBirula2006}; the Holevo bound limits the information about an encoded parameter obtainable from measurements \cite{Holevo1973,Hall2018}. Whereas the conventional angle--angular-momentum relation \cite{BialynickiBirula2006} pairs a physical angle with angular momentum, we formulate a phase--Fourier entropic uncertainty relation for the intensity-weighted distribution of doubled modal phases, with the Fourier index labeling phase harmonics. Relative entropies of rotational asymmetry, developed for quantum reference frames and metrology \cite{Hall2018,Gour2009,Marvian2014}, quantify distinguishability under phase rotations at quantum and classical levels; we bring them to non-Hermitian eigenfunctions.

In this Letter we resolve eigenfunction phase organization beyond nonorthogonality by comparing quantum and classical rotational asymmetry, linked by entropic complementarity and the Holevo bound. Because EP self-orthogonality removes only the first phase harmonic, rotational asymmetry can survive in higher orders. We show that, in classical microcavity and quantum hydrogen resonances, one branch gains this asymmetry as the other loses it near the EP despite common rigidity suppression.

For a reciprocal complex-symmetric non-Hermitian Hamiltonian, $H^T=H$, matched eigenvectors obey $H|R\rangle=E|R\rangle$ and $H^\dagger|L\rangle=E^*|L\rangle$, with $|L\rangle=|R\rangle^*$. Write $\psi=|\psi|e^{i\phi}$ and $w=|\psi|^2/\int|\psi|^2\dd\mu$ in a positive measure consistent with the overlap. Phase rigidity is then
\begin{equation}
r=\frac{|\langle L|R\rangle|}{\|L\|\|R\|}
 =\frac{|\int\psi^2\dd\mu|}{\int|\psi|^2\dd\mu}
 =\big|\langle e^{2i\phi}\rangle_w\big|.
\label{eq:rigidity}
\end{equation}
Here $\langle X\rangle_w=\int wX\dd\mu$. Doubling maps the phase-locked values $0$ and $\pi$ to the same unit phasor, giving $r=1$. The matched Petermann factor is $K=r^{-2}$ \cite{Wiersig2023}.

With periodic delta $\delta_{2\pi}$, the intensity-weighted distribution of $\theta=2\phi\bmod 2\pi$ is
\begin{equation}
p(\theta)=\int w(x)\delta_{2\pi}[\theta-2\phi(x)]\dd\mu(x).
\label{eq:marginal}
\end{equation}
Its circular moments are $Z_k=\int_0^{2\pi}p(\theta)e^{ik\theta}\dd\theta$ \cite{MardiaJupp1999}, so $r=|Z_1|$. Global rephasing by $\beta_0$ rotates $p$ by $\alpha=2\beta_0$ and sends $Z_k\mapsto e^{ik\alpha}Z_k$. For an isolated nondegenerate reciprocal scalar mode, identical weak probes in a homogeneous region give $\delta\omega(x)=C\psi(x)^2$ with a common complex constant $C$ \cite{Cognee2019}. The $|\delta\omega|$-weighted shift directions recover $p$ up to $\arg C$ (Supplemental Sec.~C.1 \cite{Supplement}). Both offsets rotate $p$ by $U(1)$, preserving moment magnitudes and continuous entropies. Full $U(1)$ invariance of $p$ itself requires $Z_k=0$ for every $k\ne0$, equivalently $p=(2\pi)^{-1}$. At an EP, self-orthogonality enforces only $Z_1=0$, so vanishing rigidity need not restore rotational symmetry.

For a phase density $p$, the square-root amplitude $s=\sqrt p$ defines a normalized auxiliary state \cite{Wootters1981},
\begin{equation}
|s\rangle=\int_0^{2\pi}\sqrt{p(\theta)}\,|\theta\rangle\dd\theta,
\quad \langle s|s\rangle=1.
\label{eq:encoding}
\end{equation}
The phase basis obeys $\langle\theta|\theta'\rangle=\delta_{2\pi}(\theta-\theta')$. The encoding is auxiliary to the physical resonance. Encoding overlaps give Bhattacharyya coefficients \cite{FuchsVandeGraaf1999}, $\langle s|s'\rangle=\int_0^{2\pi}\sqrt{pp'}\,\dd\theta=1-d_H^2(p,p')$, with Hellinger distance $d_H$. The Fourier amplitudes of $s$ are $c_m=(2\pi)^{-1/2}\int_0^{2\pi}\sqrt p\,e^{-im\theta}\dd\theta$, with $q_m=|c_m|^2$ and $\sum_m q_m=1$. The complementary entropies are
\begin{equation}
h_\phi=-\int_0^{2\pi}p\log_2p\,\dd\theta,
\quad H_F=-\sum_mq_m\log_2q_m.
\label{eq:entropies}
\end{equation}
Their continuous phase--Fourier uncertainty relation is \cite{BialynickiBirula2006}
\begin{equation}
U\equiv h_\phi+H_F\ge\log_2(2\pi).
\label{eq:EUR}
\end{equation}
For $M=360$ equal bins \cite{Supplement}, $H_\phi^{(M)}$ is the Shannon entropy of the bin probabilities and $H_F^{(M)}$ that of the squared unitary Fourier amplitudes of their square roots. The Maassen--Uffink bound is $U_M=H_\phi^{(M)}+H_F^{(M)}\ge\log_2 M$ \cite{MaassenUffink1988}. We omit bin superscripts and write $U=U_M$.

\begin{figure}[t]
\centering
\includegraphics[width=\columnwidth]{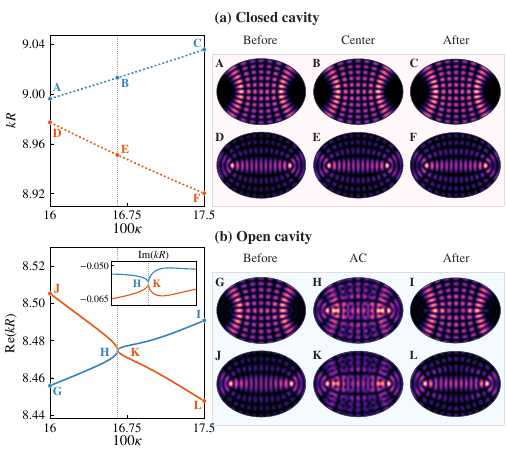}
\caption{Computed spectra and intensity patterns in the elliptic cavity. (a) Closed-cavity spectrum and modes A--F. (b) Open-cavity spectrum on the broad cut ($n=2.97000$). Blue and orange denote modes \textbf{a} (A--C, G--I) and \textbf{b} (D--F, J--L), respectively. The inset shows $\operatorname{Im}(kR)$. The dotted guide marks the sampled interaction center (AC), $100\kappa=16.655$. Here $R=1$ in cavity units. Maps show relative intensities.}
\label{fig:cavity_modes}
\end{figure}

\begin{figure}[t]
\centering
\includegraphics[width=\columnwidth]{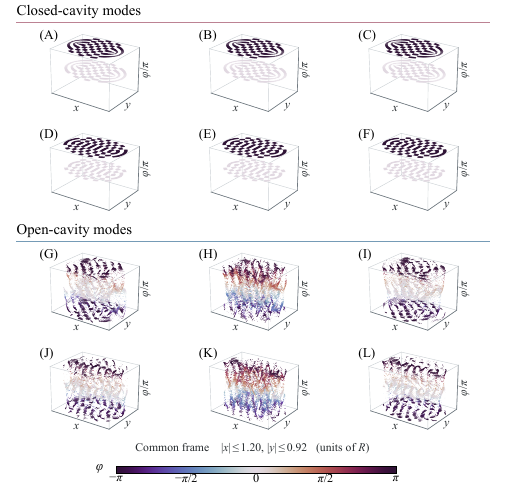}
\caption{Phase maps of the modes whose intensities appear in Fig.~\ref{fig:cavity_modes}, with matching labels A--L. Closed-cavity modes (A--F) have local phases $0$ or $\pi$. For all panels, height gives $\phi/\pi$ and color gives $\phi$ in radians.}
\label{fig:phase_maps}
\end{figure}

Optical microcavities host EPs \cite{Lee2009Microcavity,Park2020MaximalShannon}. The elliptic cavity has $x^2/a^2+y^2/b^2=1$, $a=1+\kappa$,
$b=(1+\kappa)^{-1}$, with lengths in units of the equal-area radius $R$. Outgoing TM resonances, $\psi=E_z$, satisfy
$[\nabla_{x,y}^2+n^2(x,y)(kR)^2]\psi=0$, with index $n$ inside and unity
outside. Boundary-element solutions \cite{Wiersig2003BEM} use broad and
near-EP cuts (Supplemental Material \cite{Supplement}).
The interior observable
$\rOm=|\int_\Omega\psi^2\dd\mu|/\int_\Omega|\psi|^2\dd\mu$
is distinct from full outgoing-resonance normalization. Similar intensities [Fig.~\ref{fig:cavity_modes}] can conceal broad phase distributions [Fig.~\ref{fig:phase_maps}].

\begin{figure*}[t]
\centering
\includegraphics[width=\textwidth]{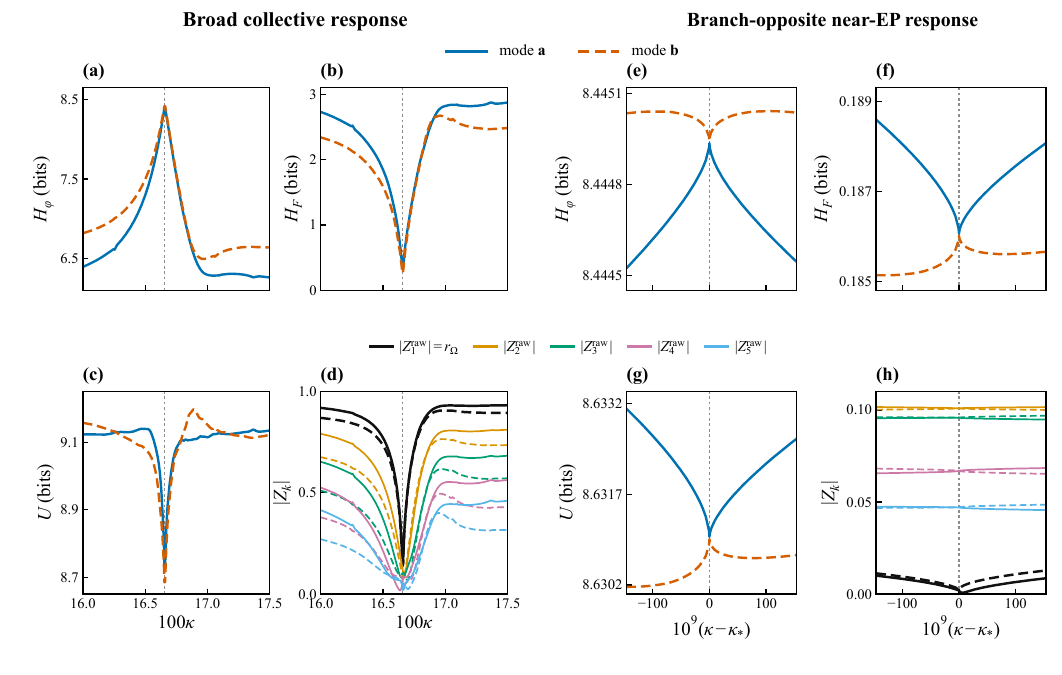}
\caption{Cavity entropies $H_\phi$, $H_F$, and $U$ in bits, in (a,e), (b,f), and (c,g). Panels (a)--(d) show the broad cut of Figs.~\ref{fig:cavity_modes} and \ref{fig:phase_maps} ($n=2.97000$), and (e)--(h) show the near-EP cut ($n=2.97722367$). Dotted markers give $\kappa=0.16655$ and the near-cut origin $\kappa_*=0.166571791302$. Blue solid and orange dashed entropy curves denote \textbf{a} and \textbf{b}. Panels (d,h) plot $|Z_k^{\rm raw}|$, $k=1,\ldots,5$, with color denoting order and line style denoting branch. $|Z_1^{\rm raw}|=\rOm$ \cite{Supplement}.}
\label{fig:cavity_entropies}
\end{figure*}

On the broad cut, both phase distributions spread as rigidity falls, with peaks in $H_\phi$ and dips in $H_F$ [Fig.~\ref{fig:cavity_entropies}(a,b,d)]. Near the EP, one distribution broadens while the other sharpens, with opposite Fourier-entropy changes despite rigidity dips on both modes [Fig.~\ref{fig:cavity_entropies}(e,f,h)].

The circular moments resolve these responses,
\begin{equation}
p(\theta)=\frac{1}{2\pi}\Big[1+2\sum_{k\ge1}|Z_k|\cos(k\theta-\arg Z_k)\Big].
\label{eq:moments}
\end{equation}
The same moments defined after Eq.~\eqref{eq:marginal} satisfy $Z_k=\sum_m c_mc_{m+k}^*$. Each $Z_k$ determines the order-$k$ harmonic of $p$ and acquires the phase factor $e^{ik\alpha}$ under rotation \cite{Marvian2014}. Rigidity $r=|Z_1|$ retains only the first harmonic's magnitude. Cancellation of $Z_1=\sum_m c_mc_{m+1}^*$, a sum over adjacent Fourier pairs, need not erase the encoding's individual Fourier coherences.

The Pearson divergence $\chi^2$ measures the mean squared relative deviation from the uniform density. For $p\in L^2$, Parseval expresses it in the harmonic powers:
\begingroup\small\thinmuskip=1mu \medmuskip=1mu \thickmuskip=1mu
\begin{equation}
\chi^2\bigl(p\Vert(2\pi)^{-1}\bigr)=2\pi\!\int_0^{2\pi}\![p-(2\pi)^{-1}]^2\dd\theta
=2\bigl(|Z_1|^2+S_{\ge2}\bigr).
\label{eq:power}
\end{equation}
\endgroup
Here $S_{\ge2}=\sum_{k\ge2}|Z_k|^2$. The entropy deficit satisfies $\log_2(2\pi)-h_\phi\le\log_2(1+\chi^2)$ and, for weak patterns, $\log_2(2\pi)-h_\phi\simeq(|Z_1|^2+S_{\ge2})/\ln 2$. On the broad cut, first and higher moments fall together [Fig.~\ref{fig:cavity_entropies}(d)], consistent with phase spreading. Near the EP, the first moment is strongly suppressed while higher moments persist [Fig.~\ref{fig:cavity_entropies}(h)], allowing the distribution to remain nonuniform even as rigidity vanishes.

Along a regular approach to a second-order EP (EP2), write $p_\pm=p_0\pm\varepsilon g+o(\varepsilon)$, with coalesced density $p_0$, branch-odd variation $g$ satisfying $\int g\,\dd\theta=0$, and real square-root coordinate $\varepsilon>0$. At the EP, self-orthogonality gives $Z_1[p_0]=0$, so both branches have the same first-order rigidity, $|Z_1[p_\pm]|=\varepsilon|Z_1[g]|+o(\varepsilon)$. For a positive reference density admitting entropy differentiation, the slope is
\begingroup\thinmuskip=1mu \medmuskip=1mu \thickmuskip=1mu
\begin{equation}
\left.\frac{\dd h_\phi[p_\pm]}{\dd\varepsilon}\right|_{0^+}
=\mp\underbrace{\int_0^{2\pi}\!g\log_2(2\pi p_0)\,\dd\theta}_{L_D}
=\mp2\operatorname{Re}\!\sum_{k\ge1}G_k\ell_k^*.
\label{eq:tangent}
\end{equation}
\endgroup
Here $G_k=Z_k[g]$ are the Fourier coefficients of the density variation and $\ell_k=(2\pi)^{-1}\int_0^{2\pi}\log_2(2\pi p_0)e^{ik\theta}\dd\theta$ those of the log-density ratio. The slope is $\mp(L_D^{(1)}+L_D^{(\ge2)})$, with $L_D^{(1)}$ from $k=1$ and $L_D^{(\ge2)}$ from higher orders. Nonzero $L_D$ sets the strength and sign of the opposite slopes. A uniform $p_0$ gives $L_D=0$, so a nonzero leading split at zero rigidity requires surviving higher harmonics in $p_0$.

\begin{figure}[!t]
\centering
\includegraphics[width=\columnwidth]{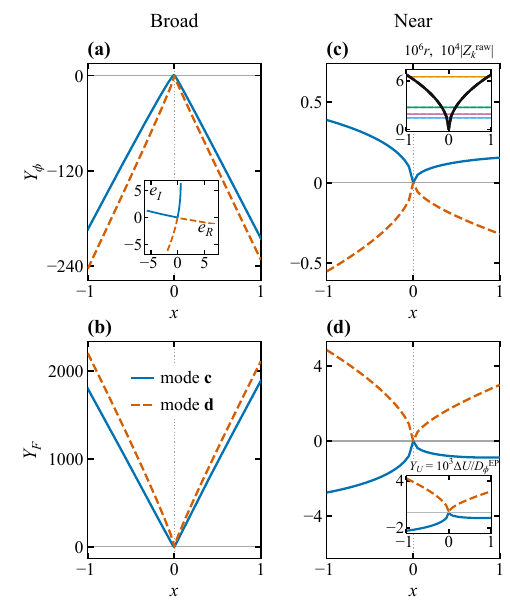}
\input{figure4_source/Fig4_caption.tex}
\end{figure}

These entropy responses also have a symmetry interpretation. The phase-entropy deficit and Fourier entropy quantify the $U(1)$ rotational asymmetry of the doubled-phase distribution and its auxiliary quantum encoding, respectively. Define the phase deficit from uniformity,
\begin{equation}
D_\phi=D\bigl(p\Vert(2\pi)^{-1}\bigr)=\log_2(2\pi)-h_\phi.
\label{eq:phase_deficit}
\end{equation}
Rotations $T_\alpha=e^{-i\alpha\hat m}$, with $\hat m|m\rangle=m|m\rangle$, shift the encoding's phase origin, $(T_\alpha s)(\theta)=s(\theta-\alpha)$, independently of any dynamical symmetry of the Hamiltonian.

Following the reference-frame construction \cite{Gour2009,Marvian2014}, we record a uniform rotation $\alpha\in[0,2\pi)$ in a classical register $A$ and place the rotated encoding in $Q$, giving
\begin{equation}
\rho_{AQ}=\int_0^{2\pi}\frac{\dd\alpha}{2\pi}\,
|\alpha\rangle\langle\alpha|\otimes T_\alpha\rho_sT_\alpha^\dagger,
\label{eq:twirl}
\end{equation}
where $\rho_s=|s\rangle\langle s|$. Discarding $A$ gives the twirl $\mathcal G(\rho_s)=\sum_mq_m|m\rangle\langle m|$, which removes Fourier coherences. Angle readout $\Theta$ yields $p_\alpha(\theta)=p(\theta-\alpha)$ with uniform marginal $\mathcal G(p)=(2\pi)^{-1}$. Correlations with $A$ give
\begingroup
\small\thinmuskip=1mu \medmuskip=1mu \thickmuskip=1mu
\begin{equation}
H_F=I(A{:}Q)
=\int_0^{2\pi}\!\frac{\dd\alpha}{2\pi}D(T_\alpha\rho_sT_\alpha^\dagger\Vert\mathcal G(\rho_s))
=D(\rho_s\Vert\mathcal G(\rho_s)).
\label{eq:asymmetry}
\end{equation}
\begin{equation}
D_\phi=I(A{:}\Theta)
=\int_0^{2\pi}\!\frac{\dd\alpha}{2\pi}D(p_\alpha\Vert\mathcal G(p))
=D(p\Vert(2\pi)^{-1}).
\label{eq:classical_asymmetry}
\end{equation}
\endgroup
Here $I$ is mutual information and $S$ is von Neumann entropy. $H_F$ quantifies the quantum resource of $U(1)$ asymmetry in the square-root encoding $\rho_s$ \cite{Gour2009}. The final equalities use rotation-invariant twirls. For pure conditional encodings, $H_F=S(\mathcal G(\rho_s))$ is the ensemble's Holevo information \cite{Hall2018} and the relative entropy of Fourier-basis coherence of $\rho_s$ \cite{Baumgratz2014}. Since $h(\Theta)=\log_2(2\pi)$ and $h(\Theta|A)=h_\phi$, $D_\phi$ is the information retained in one angle record.

Angle readout maps $\rho_s$ to $p$ and $\mathcal G(\rho_s)$ to the uniform density, since each Fourier state has a uniform angle distribution. Relative-entropy contraction \cite{Lindblad1975} gives
\begin{equation}
D\bigl(p\Vert(2\pi)^{-1}\bigr)
\le D\bigl(\rho_s\Vert\mathcal G(\rho_s)\bigr).
\label{eq:contraction}
\end{equation}
The Holevo bound on angle information thus takes the form of the phase--Fourier uncertainty relation \cite{Hall2018}. Its gap,
\begin{equation}
\mathcal E\equiv I(A{:}Q)-I(A{:}\Theta)
=h_\phi+H_F-\log_2(2\pi)\ge0,
\label{eq:gap}
\end{equation}
is the excess of quantum over classical rotational asymmetry. It bounds any additional information beyond angle readout for a single copy of the encoding. For $M$ bins and uniform cyclic shifts, $\mathcal E=U_M-\log_2M$.

Near the cavity EP, classical ($D_\phi=\log_2M-H_\phi$) and quantum ($H_F$) asymmetries decrease on \textbf{a} but increase on \textbf{b} as rigidity falls on both [Fig.~\ref{fig:cavity_entropies}(e,f,h)]. For response-weighted angle records independent at fixed probe offset $\zeta$ ($\arg C$ for simple-mode shifts), let $A_n$ be the dimensionless reduction in optimal mean circular loss $1-\cos(\hat\zeta-\zeta)$ from unit risk after $n$ records under a uniform prior \cite{Supplement}. One record gives $A_1=|Z_1|=r$ for ideal readout. Two give $A_2\ge\max_k|Z_kZ_{k+1}|$, which can remain positive at $Z_1=0$ if adjacent higher harmonics survive. At the smallest sampled cavity gap, both modes give $D_\phi\simeq0.0469$ bits per angle and resolved (angle-smoothed) $|Z_1|\simeq0.002$. The predicted $A_2=0.0153$ for each exceeds $0.00270$--$0.00294$ for first-harmonic references matched in resolved rigidity (Supplemental Sec.~C.2 and Table~S6 \cite{Supplement}).

In this window, the phase distributions are nearly uniform [Fig.~\ref{fig:cavity_entropies}(e)] and hence nearly $U(1)$-invariant. To resolve their residual asymmetry, separate the encoding into its constant component $e_0=(2\pi)^{-1/2}$ and the remainder,
\begin{equation}
s=\sqrt{1-\eta}\,e_0+\sqrt\eta\,s_\perp,
\label{eq:split}
\end{equation}
with $\eta=1-q_0$ and $s_\perp$ normalized and orthogonal to $e_0$. The nonconstant Fourier weight $\eta$ equals the rotation-averaged squared Hellinger distance to shifted copies of $p$ (Supplemental Sec.~A.4 \cite{Supplement}). For $\eta>0$, set $\nu_m=q_m/\eta$ for $m\ne0$. Then $H_F=h_2(\eta)+\eta H(\nu)$, with binary entropy $h_2$ and nonconstant-sector entropy $H(\nu)$. Thus
\begin{equation}
\dd H_F=\left[\log_2\frac{1-\eta}{\eta}+H(\nu)\right]\dd\eta
+\eta\,\dd H(\nu).
\label{eq:HFderivative}
\end{equation}
For a fixed smooth zero-mean pattern $g_0\ne0$, take $p_t=(1+tg_0)/(2\pi)$. As $t\to0$, $D_\phi\simeq2\eta/\ln2$ while $H_F\ge h_2(\eta)\sim\eta\log_2(1/\eta)$. The logarithm amplifies weight transfer, and redistribution enters with prefactor $\eta$. Thus $H_F/D_\phi\to\infty$, and sufficiently weak nonzero patterns obey $\mathcal E\ge D_\phi\simeq(r^2+S_{\ge2})/\ln2$ (Supplemental Sec.~A.4 \cite{Supplement}). Both remain positive at $r=0$ if weak higher harmonics survive. At the smallest sampled gap, both resolved cavity distributions give $\mathcal E\simeq0.139$ bits, about $3D_\phi$ (Table~S6 \cite{Supplement}).

For a strength-only response, use $p_\lambda=(1-\lambda)/(2\pi)+\lambda p>0$, which rescales nonzero harmonics by $\lambda>0$ at fixed shape. The local ratio (primes denote $\mathrm d/\mathrm d\lambda$) is
\begin{equation}
R_{\rm null}=\frac{H_F'(1)}{D_\phi'(1)}
\simeq\frac{2H_F-D_\phi}{2D_\phi}=\frac12+\frac{\mathcal E}{D_\phi}.
\label{eq:Rnull}
\end{equation}
The denominator is $D(p\Vert(2\pi)^{-1})+D((2\pi)^{-1}\Vert p)$. For weak fixed patterns, $D_\phi'(1)\simeq2D_\phi$ and $H_F'(1)\simeq2H_F-D_\phi$. The approximation agrees with the exact cavity values $3.46$--$3.51$ within $0.7\%$ (Supplemental Sec.~A.5 \cite{Supplement}).

Each finite baseline starts from the observed initial distribution and matches the final $D_\phi$. The exact budget is
\begin{equation}
\Delta U=W+C_{\rm int}+\Delta H_\phi,
\label{eq:budget}
\end{equation}
where $\Delta$ denotes final minus initial on each branch and bars denote endpoint means. Weight transfer between constant and nonconstant components gives $W=\Delta h_2+\overline{H(\nu)}\,\Delta\eta$. Redistribution among nonconstant orders gives $C_{\rm int}=\bar\eta\,\Delta H(\nu)$. Across the four near-EP approaches, $|W|>|C_{\rm int}|+|\Delta H_\phi|$, and the two asymmetries change in the same direction. The broad cut shares this sign relation \cite{Supplement}. Three near-EP response ratios $R_{\rm obs}=|\Delta H_F/\Delta D_\phi|=4.98$--$9.27$ exceed their matched finite ratios $R_0=3.47$--$3.51$, signaling harmonic reorganization. On the right of \textbf{b}, $R_{\rm obs}=3.36$ lies close to $R_0=3.48$, while opposing residuals in $W$ and $C_{\rm int}$ expose reorganization (Table~S1 and Supplemental Sec.~A.6 \cite{Supplement}).

We test the same organization in hydrogen in parallel electric and magnetic fields \cite{Feldmaier2016}. Complex scaling \cite{SMMoiseyev1998} gives the atomic-unit Hamiltonian
\begin{equation}
H_\vartheta=\tfrac12e^{-2i\vartheta}\mathbf p^2-\tfrac{e^{-i\vartheta}}\varrho
+\tfrac\gamma2L_z+\tfrac{\gamma^2e^{2i\vartheta}}8\varrho_\perp^2+fe^{i\vartheta}z.
\label{eq:hydrogen}
\end{equation}
Here $\gamma,f$ are magnetic and electric field strengths, respectively, $\mathbf p$ is electron momentum, $\varrho=|\mathbf r|$, $\varrho_\perp^2=x^2+y^2$, and $\vartheta$ is the scaling angle. We take $L_z=0$ (Supplemental Part~B \cite{Supplement}).

Along the hydrogen EP path [Fig.~\ref{fig:hydrogen}], the cavity behavior recurs: both modes share broad phase-entropy peaks and Fourier-entropy dips and lose rigidity, yet near the EP one gains rotational asymmetry as the other loses it. Quantum asymmetry again changes more than classical asymmetry in endpoint-to-EP comparisons (Supplemental Part~B \cite{Supplement}).

In conclusion, EP self-orthogonality forces only the first phase harmonic to vanish, so modes with nearly equal rigidity can display opposite rotational-asymmetry responses. At isolated reciprocal scalar EP2s, higher harmonics predict preferred probe-induced splitting axes in the complex-frequency plane (Supplemental Sec.~C.1 and Fig.~S1 \cite{Supplement}, see also Ref.~\cite{Wiersig2022Response}). With calibrated local probes, these axes set the balance between frequency and linewidth splitting, which matters for EP sensing \cite{Wiersig2014Sensing}. For normalized pure modes in a fixed real basis, geometric imaginarity is $\mathcal I_g=(1-r)/2$ \cite{Wu2021Quantification,Wu2021Operational}. Higher harmonics distinguish encoded $U(1)$ asymmetries at fixed $\mathcal I_g$ (Supplemental Sec.~A.3 \cite{Supplement}). Rotational asymmetry thus connects phase organization missed by rigidity to measurable resonance responses and encoded quantum resources.

\begin{acknowledgments}
This work was supported by the National Research Foundation of Korea (NRF) through a grant funded by the Ministry of Science and ICT (Grant Nos. RS-2023-00211817 and RS-2025-00515537). This work was also supported by the Institute for Information \& Communications Technology Promotion (IITP) grant funded by the Korean government (MSIP) (Grant Nos. RS-2025-02304540 and RS-2019-II190003), the National Research Council of Science \& Technology (NST) (Grant No. GTL25011-000), and the Korea Institute of Science and Technology (KISTI) (Grant No. P26028). S.L. acknowledges support from the National Research Foundation of Korea (NRF) grants funded by the MSIT (Grant No. RS-2022-NR068791). We acknowledge the Yonsei University Quantum Computing Project Group for its support. This work was supported using HPC resources and technical support provided by KISTI.
\end{acknowledgments}

\textit{Data availability.} Numerical tables and computational details are provided in the Supplemental Material \cite{Supplement}. Readers may contact the corresponding author about access to the underlying simulation data and analysis scripts.

\onecolumngrid
\smpage{1}
\smpage{2}
\smpage{3}
\smpage{4}
\smpage{5}
\smpage{6}
\smpage{7}
\smpage{8}
\smpage{9}
\smpage{10}
\smpage{11}
\clearpage

\end{document}